\documentclass[aps, prd, onecolumn, nofootinbib, floatfix, showpacs,
superscriptaddress]{revtex4-2}
\usepackage{subfigure}
\usepackage{epsfig}
\usepackage{amsmath}
\usepackage{amsfonts}
\usepackage{amssymb}
\usepackage{amssymb,amsmath,amsfonts}
\usepackage{xfrac}
\usepackage{color}
\usepackage[utf8]{inputenc}
\usepackage{graphicx}
\usepackage{dcolumn}
\usepackage{bm}
\usepackage{tikz}

\usepackage{tcolorbox}
\usepackage{hyperref}
\hypersetup{colorlinks, citecolor=red, linkcolor=bluscuro, urlcolor=bluscuro}
\definecolor{rossos}{cmyk}{0,1,1,0.55}
\definecolor{bluscuro}{rgb}{0.15, 0.2, .85}
\definecolor{bluchiaro}{cmyk}{1,.3,0.,0.1}

\usepackage{verbatim}

\newcommand{\de}{\mathrm{d}}
\newcommand{\lin}{\\[7pt]}
 
\newcommand{\beqa}{\begin{eqnarray}}
\newcommand{\eeqa}{\end{eqnarray}}

\newcommand{\pder}[2]{\dfrac{\partial#1}{\partial#2}}
\newcommand{\pdder}[3]{\dfrac{\partial^2 #1}{\partial #2 \partial #3}}

\newcommand{\dder}[2]{\dfrac{\delta#1}{\delta#2}}
\newcommand{\pdot}[1]{\dot{\partial}_{#1}}

\newcommand{\D}{\mathcal{D}}
\newcommand{\Gd}{\mathcal{G}}

\newcommand{\R}{\mathcal{R}}

\begin{document}

\title{  Finsler-Randers-Sasaki gravity and cosmology}

\author{E. Kapsabelis}
\affiliation{Department of Physics, National and Kapodistrian University of 
Athens, 	Panepistimiopolis 15784, Athens, Greece}

\author{Emmanuel N. Saridakis}
\affiliation{National Observatory of Athens, Lofos Nymfon, 11852 Athens, Greece}
\affiliation{CAS Key Laboratory for Research in Galaxies and Cosmology,  
 University of Science and Technology of China, Hefei, Anhui 230026, China}
\affiliation{Departamento de Matem\'{a}ticas, Universidad Cat\'{o}lica del 
Norte, Avda.
Angamos 0610, Casilla 1280 Antofagasta, Chile}

\author{P. C. Stavrinos}
\affiliation{Department of Mathematics, National and Kapodistrian University of 
Athens,
	Panepistimiopolis 15784, Athens, Greece}

\begin{abstract} 
We present for the first time a Friedmann-like construction in the 
framework of an osculating   Finsler-Randers-Sasaki  geometry. 
In 
particular, 
we consider a vector field in the metric on a Lorentz tangent bundle, and thus 
the curvatures of horizontal and vertical spaces, as well as the extra 
contributions of torsion and non-linear connection, provide an 
intrinsic richer geometrical structure, with additional degrees of freedom, 
that 
lead to extra terms in the field equations. Applying these modified field 
equations at a cosmological setup  we extract the generalized Friedmann 
equations for the horizontal and vertical space, showing that we obtain 
an effective dark energy sector arising from 
the richer underlying structure of the tangent bundle. Additionally, as it is 
common in Finsler-like constructions, we obtain an effective interaction 
between 
matter and   geometry. Finally, we consider a specific model  and we show that 
it can describe the sequence of 
matter and dark-energy epochs, and that the dark-energy 
equation of state can lie in the quintessence or phantom regimes, or cross 
the phantom divide. 
             
\end{abstract}

\pacs{98.80.-k, 95.36.+x, 04.50.Kd}

\maketitle

\section{Introduction}

General Relativity (GR) has been proved a successful theory of gravity, tested 
with high precision at Earth-based  and Solar System experiments (perihelion  
preccession  of Mercury, gravitational  redshift, Shapiro time-delay effect, 
etc)
\cite{Will:2014kxa}. Nevertheless, at the theoretical level one faces the 
problem of non-renormalizability \cite{Addazi:2021xuf}, since GR cannot be 
incorporated in a quantum description \cite{AlvesBatista:2023wqm}. 
Additionally, at the cosmological level we still have the open issues 
of dark matter and dark energy \cite{Copeland:2006wr,Cai:2009zp}, as well as 
possible tensions between predictions and observations, such as the $H_0$ 
\cite{DiValentino:2020zio} and 
$\sigma_8$ tensions \cite{DiValentino:2020vvd} (for a review see 
\cite{Abdalla:2022yfr}).
Hence, a large amount of research was devoted in the construction of various 
gravitational modifications, namely theories that possess general relativity at 
as a particular limit, but which in general exhibit richer behavior, 
theoretically and cosmologically improved  
\cite{CANTATA:2021ktz,Capozziello:2011et,Cai:2015emx,
Nojiri:2017ncd}.

The basic procedures towards modified gravity constructions is to start from 
the  Einstein-Hilbert  Lagrangian and add   extra terms, resulting to    
 $f(R)$ gravity  \cite{Starobinsky:1980te,Capozziello:2002rd},   $f(G)$
gravity \cite{Nojiri:2005jg},    Weyl gravity \cite{Mannheim:1988dj},   
  Lovelock gravity \cite{Lovelock:1971yv}, etc. Furthermore, one  can     
consider  alternative geometries, beyond the Riemannian one, such is the 
torsional formulation of gravity, and construct extensions such as   
$f(T)$ gravity \cite{Cai:2015emx}, $f(T,T_G)$ gravity
\cite{Kofinas:2014owa}, $f(T,B)$ gravity, etc. In similar lines, one can use
 non-metricity, resulting to symmetric teleparallel gravity  
\cite{Bahamonde:2015zma,BeltranJimenez:2017tkd}, $f(Q)$ 
gravity \cite{Anagnostopoulos:2021ydo}, etc.

However, one can proceed to more radical geometrical modifications, namely use 
Finsler and Finsler-like geometries, which have richer structure than 
Riemannian framework, and use them in order to construct gravitational theories 
\cite{Miron,Vacaru:2005ht,stavrinos-ikeda1999, Vacaru:2002,
stavrinos-ikeda2000,kour-stath-st2012,Triantafyllopoulos:2018bli, 
Minas:2019urp,Konitopoulos:2021eav,
Kouretsis:2010vs,Mavromatos:2010jt,stavrinos-alexiou,Papagiannopoulos:2017whb,
Basilakos:2013hua, 
Basilakos:2013ij,Vacaru:2010fi,Vacaru:2010rd,Kostelecky:2008be,
Mavromatos:2010nk,Chang:2007vq,
Foster:2015yta,Kostelecky:2011qz,Pfeifer:2011xi,Kostelecky:2012ac,
Stavrinos:2012ty,Stavrinos:2021ygh,
Hohmann:2016pyt,Hohmann:2018rpp,Pfeifer:2019wus,
Papagiannopoulos:2020mmm,Hohmann:2020mgs,
Ikeda:2019ckp}.
These modified theories of gravity use generalized metric structures, where a 
vector field is incorporated in the geometrical construction, and have 
contributed
with different directions in the development of locally anisotropic models for 
the gravitational field theory and   cosmology.

 In Finsler and Finsler-like geometries more than one connection and curvature 
appear, which depend on the position and velocity, in contrast to GR in which 
there is only the Levi-Civita connection and the curvature of the Riemannian 
space. Therefore,   gravity  can be studied in a different way in the framework 
of an 8-dimensional Lorentz tangent bundle or a vector bundle which includes 
the observer (velocity/tangent vector) with extra internal/dynamical degrees of 
freedom 
\cite{Vacaru:2010fi,Stavrinos:2012ty,Hohmann:2018rpp,
Triantafyllopoulos:2018bli,Minas:2019urp,Triantafyllopoulos:2020ogl,
Konitopoulos:2021eav}, as well as in an oscullating Riemannian and Barthel 
framework \cite{Hama:2023bkl,Bouali:2023flv,Hama:2022vob}. 

Concerning this 
approach, all kinds of generalized metric theories belong to the larger class 
of the so called ``anisotropic field theories'', since Lorentz violations, 
velocity fields and torsions produce anisotropies in the 
space and the matter sector 
\cite{Stavrinos:2006rf,Kouretsis:2008ha,Kostelecky:2011qz,Kostelecky:2012ac}. 
Hence, these internal geometrical anisotropies, which should not be confused 
with spacetime anisotropies that may appear in Riemannian geometry too (e.g. 
in Bianchi cases)  are induced by internal direction/velocity $y$-variables in 
addition to the position $x$-variables in the structure of the base manifold. 
In these lines,     geometrical anisotropies   can be   considered as a 
``potential'' or a tidal field in the matter sector \cite{wheeler1973}. In 
cases 
where an anisotropy is included in the metric structure of spacetime, as it 
appears in Finsler and Finsler-like cosmologies, it is incorporated in the 
effective energy-momentum tensor of the anisotropic structure, which could 
potentially lead to energy exchange between  geometry and matter    
\cite{Hama:2021frk}. Finally, similarly to general relativity, geometrical 
effects are produced not only by the distribution of mass-energy but also by 
its 
motion \cite{Weinberg:1972kfs}.

In this work we propose a novel geometrical structure, namely that of 
  Finsler-Randers-Sasaki  (F-R-S)  type, in order to extract   generalized 
gravitational field 
equations. Then, applying them in a cosmological framework we construct 
   F-R-S  cosmology, which is characterized by modified 
Friedmann 
equations with new terms that depend on the underlying geometry and the  
tangent bundle features. As we will show, these terms can lead to interesting 
cosmological implications, and describe the thermal history of the Universe, as 
well as the effective dark energy sector. The plan of the work is the 
following: In Section \ref{Model} we provide the basic   concepts of 
     F-R-S   geometry, and we extract the general 
gravitational field 
equations. Then, in Section \ref{Cosmologyapl} we proceed to the application at 
a cosmological framework, and we extract the modified Friedmann equations  
for the horizontal and vertical subspaces, 
investigating also specific examples. 
Finally, in Section \ref{Conclusions} we discuss our main results.

\section{ Finsler-Randers-Sasaki geometry and gravity}
\label{Model}

In this section we present the basics of  Finsler-Randers-Sasaki 
geometry and gravity. 
We will start by introducing  some geometrical aspects from   Finsler geometry 
and the oscullating Riemannian metric, and then we will use it to construct a 
gravitational theory.  

\subsection{Finsler-Randers-Sasaki geometry with oscullating Riemannian metric}

We consider an $n$-dimensional bundle $M$, as well as its tangent bundle $TM$, 
with a fibered and differentiable (smooth) metric function $F(x,y)$ with 
the following properties:
\begin{enumerate}
		\item $F$ is continuous on $TM$ and smooth on  $ \widetilde{TM}\equiv 
TM\setminus \{0\} $, namely the tangent bundle minus the null set $ \{(x,y)\in 
TM | F(x,y)=0\}$ . \label{finsler field of definition}
		\item $ F $ is positively homogeneous of first degree on its second 
argument:
		\begin{equation}
			F(x^\mu,ky^\alpha) = kF(x^\mu,y^\alpha), \qquad k>0 \label{finsler 
homogeneity}.
		\end{equation}
		\item 
  For each $x\in M$ the fundamental metric tensor :
	\begin{equation}
			g_{ij}(x,y) =\pm \dfrac{1}{2}\pdder{F^2}{y^i}{y^j} \label{finsler 
metric} 
	\end{equation}
is non-singular, with $i,j=0,1,...,n-1$.
  \end{enumerate}
A Lorentz tangent bundle $TM$ over a spacetime 4-dimensional manifold $M$ is a 
fibered 8-dimensional manifold with local coordinates $\{x^\mu,y^ a\}$, where 
the Greek indices of the spacetime variables $x$ are 
$\kappa,\lambda,\mu,\nu,\ldots = 0,\ldots,3$ and the Latin indices of the fiber 
variables $y$ are  $ a, b,\ldots,f = 0,\ldots,3$. An extended Lorentzian 
structure on $TM$
can be provided if the background manifold is equipped with a Lorentz metric 
tensor of signature $(-1,\ldots,1)$.
  As it is known, a metric following the above three properties is called  a   
Finsler metric  \cite{Miron,Vacaru:2005ht}.

Additionally, one can introduce the  oscullating Riemannian metric 
on a differentiable manifold \cite{matsumoto1993}. In particular,  this  can be 
defined by a tangent vector field $Y:U\rightarrow TU$, where $U\subset M$ is an 
open neighborhood on $M$ with the property $Y(x)\neq 0$ $\forall x\in U$. In 
such a case the metric can be defined by the relation:
  \begin{equation}\label{oscullation}
      g_{ij}(x)=g_{ij}(x,y)|_{y=Y(x)} , x\in U.
  \end{equation}
    In the following we will consider that all   non-vanishing global vector 
fields, defined on the  spacetime manifold, satisfy $M$ $y(x)=Y(x)$ 
\cite{rund2012}.
    The pair $(U,g_{ij}(x,y(x))$ is called $Y$-oscullating Riemannian metric 
associated to $(M,F)$ manifold.

As it is known, the length of a curve $c$ in a Finsler space is given by the 
  integral 
  \begin{equation}\label{length}
      l(c)=\int_{a}^{b}F(x,y)d\tau,
  \end{equation}
  with $y=\frac{dx}{d\tau}$ and $\tau$ an affine parameter along the curve.
  In a    Finsler-Randers (FR) spacetime  \cite{Randers1941} the metric 
is given by the relation 
\cite{Kapsabelis:2022bue,Triantafyllopoulos:2020vkx}:
  \begin{equation}
  \label{lagrangian2}
		F(x,dx) = \left(-a_{\mu\nu}(x)dx^{\mu}dx^{\nu}\right)^{1/2} + 
f_{\alpha}dx^{\alpha},
\end{equation}
  where $a_{\mu\nu}(x)$ is a Riemannian pseudo-metric and $f_{\alpha}$ a 
covector with $||f_{\alpha}||\ll 1$. Note that the 1-form 
$f_{\alpha}dx^{\alpha}$ can be interpreted as the ``energy'' produced by the 
anisotropic force field $f_{\alpha}$, and hence due to \eqref{length} 
and \eqref{lagrangian2} the integral $\int_{a}^{b}F(x,dx)$ represents the 
``total work'' that a particle requires in order to move along a path with 
proper time $\tau$.

We proceed by writing the  corresponding Lagrangian  function of an FR space 
with an oscullating Riemannian metric, which is
  \begin{equation}
  \label{lagrangian3}
		F(x,y(x)) = \left[-g_{\mu\nu}(x,y(x))y^{\mu}(x)y^{\nu}(x)\right]^{1/2} 
+ f_{\alpha}(x)y^{\alpha}(x),
\end{equation}
  with $||f_{\alpha}||\ll 1$. 
  We mention here that in the above expression  the second term 
$f_{\alpha}(x)y^{\alpha}(x)$ can be interpreted as the ``power'' that is 
produced due to propagation of particles through the force-field 
$f_{\alpha}(x)$. Now, from  \eqref{finsler 
metric},\eqref{oscullation} and \eqref{lagrangian3} we can extract  the metric 
tensor $v_{\alpha\beta}(x,y(x))$ as
 \begin{equation}\label{vab}
		v_{\alpha\beta}(x,y(x)) = g_{\alpha\beta}(x) + h_{\alpha\beta}(x,y(x)), 
\end{equation}
where 
\begin{align}\label{hab}
		h_{\alpha\beta}(x,y(x)) = \frac{1}{L}(A_{\beta}g_{\alpha\gamma} + 
A_{\gamma}g_{\alpha\beta} + A_{\alpha}g_{\beta\gamma})y^\gamma(x) + 
\frac{1}{L^3}A_{\gamma}g_{\alpha\epsilon}g_{\beta\delta}y^\gamma(x) y^\delta(x) 
y^\epsilon(x),
	\end{align}
	with $L = \sqrt{-g_{\alpha\beta}y^{\alpha}(x)y^{\beta}(x)}$ and 
$||A_{\gamma}||<<1$.
Due to   
relation \eqref{hab}, the metric  \eqref{vab} can be called ``weak 
Finslerian metric'' \cite{Triantafyllopoulos:2020vkx}. Hence, as  we can see 
from \eqref{vab}, the 
term $h_{\alpha\beta}(x,y)$ can be considered as a perturbation.  
For  convenience, and in order to make notation lighter, in the following we 
will write $y$ instead of $y(x)$.

Let us now introduce the   Sasaki-type metric on $TM$   
\cite{Sasaki1958}. Such a metric 
has the form:
 \begin{equation}
		\mathcal{G} = g_{\mu\nu}(x,y)\,\mathrm{d}x^\mu \otimes \mathrm{d}x^\nu 
+ 
v_{\alpha\beta}(x,y)\,\delta y^\alpha \otimes \delta y^\beta \label{bundle 
metric}.
	\end{equation}
 In our approach we   consider that the Finslerian metric 
$v_{\alpha\beta}(x,y)$ is given by  \eqref{vab}, and the unified adapted 
frame is defined in the form
 $E_A = \,\{\delta_\mu,\dot\partial_\alpha\} $ with
\begin{equation}
\delta_\mu = \dfrac{\delta}{\delta x^\mu}= \pder{}{x^\mu} - 
N^\alpha_\mu(x,y)\pder{}{y^\alpha} \label{delta x}
\end{equation}
and
\begin{equation}
\dot \partial_\alpha = \pder{}{y^\alpha},
\end{equation}
and where $E_A$ is the adapted basis of the   tangent space $T_{x}M$.
Furthermore, we define the dual basis ${E^{A}}=(dx^{\mu},\delta y^{\alpha})$ 
with 
\begin{equation}
    \delta y^{\alpha}=dy^{\alpha}+N^{\alpha}_{\lambda}dx^{\lambda},
\end{equation}
where $E^{A}$ is the adapted basis of the cotangent bundle $T^{*}M$ and 
$N^{\alpha}_{\lambda}$ are the components of the nonlinear connection with 
$\alpha,\lambda = (0,1,2,3)$.
The nonzero coefficients of a canonical and distinguished 
$d-$connection $\mathcal D$ on $TM$ read as \cite{Miron}:
	\begin{align}
	L^\mu_{\nu\kappa} & = 
\frac{1}{2}g^{\mu\rho}\left(\delta_{\kappa}g_{\rho\nu} 
+ \delta_\nu g_{\rho\kappa} - \delta_\rho g_{\nu\kappa}\right) \label{metric 
d-connection 1}  \\
	L^\alpha_{\beta\kappa} & = \dot{\partial}_\beta N^\alpha_\kappa + 
\frac{1}{2}v^{\alpha\gamma}\left(\delta_\kappa v_{\beta\gamma} - 
v_{\delta\gamma}\,\dot{\partial}_\beta N^\delta_\kappa - 
v_{\beta\delta}\,\dot{\partial}_\gamma N^\delta_\kappa\right) \label{metric 
d-connection 2}  \\
	C^\mu_{\nu\gamma} & = \frac{1}{2}g^{\mu\rho}\dot{\partial}_\gamma 
g_{\rho\nu} \label{metric d-connection 3} \\
	C^\alpha_{\beta\gamma} & = 
\frac{1}{2}v^{\alpha\delta}\left(\dot{\partial}_\gamma v_{\delta\beta} + 
\dot{\partial}_\beta v_{\delta\gamma} - \dot{\partial}_\delta 
v_{\beta\gamma}\right) \label{metric d-connection 4}.
	\end{align}
	Finally, concerning the non-linear connection, we choose a Cartan-type of 
the form:
	\begin{equation}\label{Nconnection}
		N^\alpha_\mu = \frac{1}{2}y^\beta g^{\alpha\gamma}\partial_\mu 
g_{\beta\gamma},
	\end{equation}
which is known to have interesting applications \cite{Kapsabelis:2021dpb}.

We now have all the geometrical quantities in order to calculate the curvature 
 tensors. In particular, in such a framework the Riemann and Ricci 
curvature tensors of the horizontal space are defined as 
\cite{Miron,Vacaru:2005ht}:	
\begin{align}
	& R^\mu_{\nu\kappa\lambda} = \delta_\lambda L^\mu_{\nu\kappa} - 
\delta_\kappa L^\mu_{\nu\lambda} + L^\rho_{\nu\kappa}L^\mu_{\rho\lambda} - 
L^\rho_{\nu\lambda}L^\mu_{\rho\kappa} + 
C^\mu_{\nu\alpha}\Omega^\alpha_{\kappa\lambda} \label{R coefficients 1}\lin
	& R_{\mu\nu} = R^\kappa_{\mu\nu\kappa} =  \delta_\kappa L^\kappa_{\mu\nu} - 
\delta_\nu L^\kappa_{\mu\kappa} + L^\rho_{\mu\nu}L^\kappa_{\rho\kappa} - 
L^\rho_{\mu\kappa}L^\kappa_{\rho\nu} + 
C^\kappa_{\mu\alpha}\Omega^\alpha_{\nu\kappa} \label{d-ricci 1},
	\end{align} 
	where $\Omega^\alpha_{\nu\kappa}$ represents the curvature of the nonlinear 
connection and is defined as
\begin{equation}\label{Omega}
	\Omega^\alpha_{\nu\kappa} = \dder{N^\alpha_\nu}{x^\kappa} - 
\dder{N^\alpha_\kappa}{x^\nu}.
\end{equation} 
	Moreover, the curvature tensors of the vertical space are given by:
	\begin{align}
	S^\alpha_{\beta\gamma\delta} & = \pdot{\delta} C^\alpha_{\beta\gamma} - 
\pdot{\gamma}C^\alpha_{\beta\delta} + 
C^\epsilon_{\beta\gamma}C^\alpha_{\epsilon\delta} - 
C^\epsilon_{\beta\delta}C^\alpha_{\epsilon\gamma} \label{S coefficients 2} \lin
	S_{\alpha\beta} & = S^\gamma_{\alpha\beta\gamma} = 
\pdot{\gamma}C^\gamma_{\alpha\beta} - \pdot{\beta}C^\gamma_{\alpha\gamma} + 
C^\epsilon_{\alpha\beta}C^\gamma_{\epsilon\gamma} - 
C^\epsilon_{\alpha\gamma}C^\gamma_{\epsilon\beta} \label{d-ricci 4}.
	\end{align}
	Thus, the generalized Ricci scalar curvature in the adapted basis is:
	\begin{equation}
	\R = g^{\mu\nu}R_{\mu\nu} + v^{\alpha\beta}S_{\alpha\beta} = R+S ,
\label{bundle ricci curvature}
	\end{equation}
	where
	\begin{align}
	R=g^{\mu\nu}R_{\mu\nu} \quad,\quad
	S=v^{\alpha\beta}S_{\alpha\beta} \label{hv ricci scalar}.
	\end{align}
In the same lines, one can define the 	  torsion tensor as
	\begin{equation}\label{torsion}
		\mathcal{T}_{\nu\beta}^{\alpha} = \pdot{\beta} N_{\nu}^{\alpha} - 
L_{\beta\nu}^{\alpha},
	\end{equation}
 where $L_{\beta\nu}^{\alpha}$ is given in 
\eqref{metric d-connection 2}.

\subsection{Finsler-Randers-Sasaki gravity}

Having presented the Finsler-Randers-Sasaki geometrical framework, we 
can 
use it in 
order to construct a gravitational theory. An Einstein-Hilbert-like action  
on $TM$ 
can be defined as
\begin{equation}\label{Hilbert like action}
	K = \int_{\mathcal N} d^8\mathcal U \sqrt{|\Gd|}\, \R + 2\kappa 
\int_{\mathcal N} d^8\mathcal U \sqrt{|\Gd|}\,\mathcal L_M,
	\end{equation}
	for some closed subspace $\mathcal N\subset TM$,
	where $ L_M$ is the standard matter Lagrangian and $\kappa$ is the 
gravitational constant. Note that
\begin{equation}
	d^8\mathcal U = \de x^0 \wedge \ldots \wedge\de x^3 \wedge \de y^4 \wedge 
\ldots \wedge \de y^7,
	\end{equation}
	while the absolute value of the metric determinant  $|\Gd|$ is 
$\sqrt{|\Gd|} 
= \sqrt{-g}\sqrt{-v}$, with $g, v$ the determinants of the 
metrics $g_{\mu\nu}, v_{\alpha\beta}$ respectively, as it follows from  the 
form of \eqref{bundle metric}.	

Performing variation of the above action in terms of $g_{\mu\nu}$, 
$v_{\alpha\beta}$ and $N^\alpha_\kappa$  we extract the field 
equations  as (the details are presented in Appendix 
\ref{Fieldequations}):
 \begin{align}
		& \overline R_{\mu\nu} - \frac{1}{2}({R}+{S})\,{g_{\mu\nu}}  + 
\left(\delta^{(\lambda}_\nu\delta^{\kappa)}_\mu - g^{\kappa\lambda}g_{\mu\nu} 
\right)\left(\mathcal D_\kappa\mathcal T^\beta_{\lambda\beta} - \mathcal 
T^\gamma_{\kappa\gamma}\mathcal T^\beta_{\lambda\beta}\right)  = \kappa 
T_{\mu\nu}  \label{feq1}\\
		& S_{\alpha\beta} - \frac{1}{2}({R}+{S})\,{v_{\alpha\beta}} + 
\left(v^{\gamma\delta}v_{\alpha\beta} - 
\delta^{(\gamma}_\alpha\delta^{\delta)}_\beta \right)\left(\mathcal D_\gamma 
C^\mu_{\mu\delta} - C^\nu_{\nu\gamma}C^\mu_{\mu\delta} \right) = \kappa 
Y_{\alpha\beta} \label{feq2}\\
		& g^{\mu[\kappa}\pdot{\alpha}L^{\nu]}_{\mu\nu} +  2 \mathcal 
T^\beta_{\mu\beta}g^{\mu[\kappa}C^{\lambda]}_{\lambda\alpha} = \kappa\mathcal 
Z^\kappa_\alpha \label{feq3},
	\end{align}
 where we have defined the ``energy-momentum tensors''
	\begin{align}
		T_{\mu\nu} &\equiv - 
\frac{2}{\sqrt{|\Gd|}}\frac{\Delta\left(\sqrt{|\Gd|}\,\mathcal{L}_M\right)}{
\Delta g^{\mu\nu}} = - 
\frac{2}{\sqrt{-g}}\frac{\Delta\left(\sqrt{-g}\,\mathcal{L}_M\right)}{\Delta 
g^{\mu\nu}}\label{em1}\\
		Y_{\alpha\beta} &\equiv 
-\frac{2}{\sqrt{|\Gd|}}\frac{\Delta\left(\sqrt{|\Gd|}\,\mathcal{L}_M\right)}{
\Delta v^{\alpha\beta}}  = 
-\frac{2}{\sqrt{-v}}\frac{\Delta\left(\sqrt{-v}\,\mathcal{L}_M\right)}{\Delta 
v^{\alpha\beta}}\label{em2}\\
		\mathcal Z^\kappa_\alpha &\equiv 
-\frac{2}{\sqrt{|\Gd|}}\frac{\Delta\left(\sqrt{|\Gd|}\,\mathcal{L}_M\right)}{
\Delta N^\alpha_\kappa} = -2\frac{\Delta\mathcal{L}_M}{\Delta 
N^\alpha_\kappa}\label{em3},
	\end{align}
	where  
$\delta^\mu_\nu$ and $ \delta^\alpha_\beta$ are the Kronecker symbols.

Note that the second field equation \eqref{feq2} is simplified to:
\begin{equation}
    -\frac{1}{2}Rv_{\alpha\beta} = \kappa Y_{\alpha\beta},
    \label{feq2b}
\end{equation}
where in our case $S_{\alpha\beta}$ and $S$ are zero and the mixed Cartan 
symbols  $C^{\mu}_{\nu\gamma}$ are also zero as we can see from \eqref{metric 
d-connection 3}.
Additionally,  we mention that the third field equation \eqref{feq3} is 
identically zero since the coefficients $L^{\rho}_{\mu\nu}$ are identical to 
the 
classical to Christoffel coefficients of the FRW model, thus they do not depend 
on the vertical coordinates $y^{\alpha}$. The mixed Cartan coefficients 
$C^{\mu}_{\nu\gamma}$ are zero as we mentioned and the mixed energy-momentum 
tensor $Z^{\kappa}_{\alpha}$ is also zero from relation \eqref{em3} since we do 
not consider the matter field to depend on the non-linear connection 
$N^{\alpha}_{\kappa}$.

 Let us   make some remarks in order to give a physical interpretation of  
equations  \eqref{feq1}, \eqref{feq2} and \eqref{feq3}. As one can see, they 
may contain a source of local-matter creation  and  contribute to the 
anisotropic energy-momentum tensors $T_{\mu\nu}$ and $Y_{\alpha\beta}$  of the 
horizontal and vertical spaces. Hence, the energy-momentum tensor 
$T_{\mu\nu}$ includes  additional information of the action of the local 
anisotropy of matter fields. $Y_{\alpha\beta}$, on the other hand, is an object 
with no equivalent in Riemannian gravity, and it incorporates more 
information of intrinsic anisotropy, which is produced from the vertical metric 
structure $v_{\alpha\beta}$, and it includes additional gravitational field in
the framework of the osculating tangent bundle. Finally, the 
energy-momentum tensor $\mathcal{Z}^{\kappa}_{\alpha}$ reflects the dependence 
of matter fields on the nonlinear connection $N^{\alpha}_{\mu}$, a structure 
which induces an interaction between internal and external spaces. This tensor 
is different from $T_{\mu\nu}$ and $Y_{\alpha\beta}$, which depend on just the 
external or internal structure respectively. 

Lastly, we  introduce a new tensor, quantifying the covariant 
derivative of the 
torsion tensor, namely
\begin{equation}
\mathcal{B}_{\mu\nu}= D_{\mu}\mathcal{T}^{\beta}_{\nu\beta} 
=\delta_{\mu}\mathcal{T}^{\beta}_{\nu\beta}-L^{\kappa}_
{\mu\nu}\mathcal{T}^{\beta}_{\kappa\beta}.
\label{bteneq}
\end{equation}
Using this definition we can re-write \eqref{feq1} as:
\begin{align}
\overline R_{\mu\nu} - \frac{1}{2}({R}+{S})\,{g_{\mu\nu}}  + 
\left(\delta^{(\lambda}_\nu\delta^{\kappa)}_\mu - g^{\kappa\lambda}g_{\mu\nu} 
\right)\mathcal{B}_{\kappa\lambda}  = \kappa 
T_{\mu\nu}\Rightarrow\nonumber\\
\overline R_{\mu\nu} - \frac{1}{2}({R}+{S})\,{g_{\mu\nu}}  + 
\mathcal{B}_{(\mu\nu)}-g_{\mu\nu}\mathcal{B} = \kappa 
T_{\mu\nu},
\label{feq1B}
\end{align}
where we have omitted from the above relation the term $\mathcal 
T^\gamma_{\kappa\gamma}\mathcal T^\beta_{\lambda\beta}$ since it is of second 
order. Hence,  taking the covariant derivative of  
\eqref{feq1B} we extract the continuity 
equation for our theory, namely
\begin{align}
D^{\mu}(\mathcal{B}_{(\mu\nu)}-g_{\mu\nu}\mathcal{B}) = \kappa D^{\mu}
T_{\mu\nu}.
\label{continuityeq}
\end{align}
As expected, and as we discussed in the Introduction,  
    F-R-S  gravity, 
similarly to other Finsler-like models, gives rise to an effective interaction 
between geometry and matter, which can have interesting cosmological 
implications.\\

\section{Cosmology}
\label{Cosmologyapl}

 In this section we apply the   Finsler-Randers-Sasaki geometry and the 
 oscullating 
Riemannian framework  at a cosmological setup, and using the general field 
equations we  derive the generalized Friedmann equations. Then we will provide 
  specific examples.

\subsection{General case}

 We consider the usual  homogeneous and isotropic  
Friedmann-Robertson-Walker (FRW)   metric 
\begin{equation}    
g_{\mu\nu}(x)=diag(-1,\frac{a^{2}(t)}{1-kr^{2}},a^{2}(t)r^2,a^{2}(t)r^2sin^{2}
\theta),
\label{FRW metric1}
\end{equation}
with $a(t)$ the usual scale factor and $k=0,\pm1$ the spatial curvature,
and substituting it in 
 \eqref{bundle metric} we find:
\begin{equation}\label{full bundle metric}
G = -dt^{2}+\frac{a^{2}(t)}{1-kr^{2}}dr^{2}+a^{2}(t)r^2 
d\theta^{2}+a^{2}(t)r^2sin^{2}\theta d\phi^{2} + 
(g_{\alpha\beta}(x)+h_{\alpha\beta}(x,y))\delta y^\alpha \otimes \delta y^\beta.
\end{equation}
For simplicity, in the following we focus on the spatially-flat case 
$k=0$. 

We consider the energy momentum tensor for a perfect fluid in the 
horizontal  and the vertical space:
\begin{align}
  \label{Tmunudef}
    &T^{\mu\nu}=(\rho_m + p_m)u^{\mu}u^{\nu} + p_mg^{\mu\nu}\\
    &Y^{\alpha\beta}=(\rho_m + p_m)y^{\alpha}y^{\beta} + p_mv^{\alpha\beta},
    \label{Ymunudef}
\end{align}
where $\rho_m$ and $p_m$ are respectively the energy density and pressure of 
the matter perfect fluid, while $u^{\mu}$ and $y^{\alpha}$ are the velocities 
of 
the fluid in the horizontal  and vertical space respectively.
We notice from relations \eqref{Ymunudef} and \eqref{vab},\eqref{hab}, that the 
energy momentum tensor $Y^{\alpha\beta}$ of the vertical space constitutes an 
anisotropic perturbation of the horizontal energy momentum tensor 
$T^{\mu\nu}$.

In order to proceed, we have to consider an ansatz for $A_{\gamma}$ and 
$y^{\gamma}$. Firstly,  it proves convenient to 
introduce the following scalars:
 \begin{align}
 \label{W0choice1}
    &W_{0}=A_{0}y^{0}\\
     \label{W1choice1}
    &W_{1}=A_{1}y^{1}\\
     \label{W2choice1}
    &W_{2}=A_{2}y^{2}\\
    &W_{3}=A_{3}y^{3},
     \label{W3choice1}
 \end{align}
and as we can see,  $W_{0}$ represents the 
time-anisotropic contribution of 
our space while $W_{1}$,$W_{2}$,$W_{3}$ express the directional components of 
the anisotropic contribution. In this work we will focus on the case 
considering  $y^{2}=y^{3}=0$ in order to have only a dependance on the 
parameter $t$:
\begin{eqnarray}
\label{choice1}
 &&A_{\gamma}=(A_{0}(t),A_{1}(t),0,0)\\
 && y^{\gamma}=(y^{0},y^{1},0,0),
 \label{choice2}
\end{eqnarray}
with $y^{0}$, $y^{1}$ constants, and $A_{0}(t)$, $A_{1}(t)$ time-dependent 
functions, in agreement with  FRW symmetries.

 Inserting   the $G$-metric from  \eqref{full bundle metric} into the 
horizontal 
field equation (\ref{feq1})  of the previous section we finally extract 
the 
generalized Friedmann equations of the horizontal and vertical space 
(extracted as (\ref{fried2bapp}),(\ref{fried2bappB}) and 
(\ref{verteq1}),(\ref{verteq2}) in Appendix
\ref{friedmannequations} ) as 
  \begin{align}
\label{fried1b}     
&\left[1-\frac{5}{2L(t)}W_{1}(t)\right]\left[\frac{\dot{a}(t)}{a(t)}\right]^{2} 
- 
\frac{5}{2L(t)}[\dot{W}_{0}(t)+\dot{W}_{1}(t)]\frac{\dot{a}(t)}{a(t)}=\frac{
\kappa } {3} \rho_m(t)\\ 
&\left[1-\frac{5}{4L(t)}W_{1}(t)\right]
\frac{\ddot{a}(t)}{a(t)}+\frac{1}{2}
\left[\frac{ \dot{a} (t)}{a(t) }
\right]^{2}-\frac{5}{2L(t)}\dot{W}_{1}(t)\frac{\dot{a}(t)}{a(t)}+\frac{5}{4L(t)}
[\ddot {W}_{0}(t)
+\ddot{W}_{1}(t)]=-\frac{\kappa}{2}p_m(t),
\label{fried2b}
\end{align}
and
\begin{align}
\label{verteq1b}
&\left\{p_m(t)+\frac{1}{2\kappa}\left\{\frac{\ddot{a}(t)}{a 
(t)}+\left[\frac{\dot{a}(t)}{a(t)}\right]^{2} 
\right\}\right\}\left\{1-\frac{1}{L (t)}[3W_{0}(t)+W_{1}(t)]+\frac{1}{L(t)^{3}
}(y^{0})^{2}[W_{0}(t)+W_{1}(t)]\right\}=[\rho_m(t)+p_m(t)](y^{0})^{2},\\[7pt]
&  \left\{p_m(t)+\frac{1}{2\kappa}\left\{\frac{\ddot{a}(t)}{a 
(t)}+\left[\frac{\dot{a}(t)}{a(t)}\right]^{2} 
\right\}\right\}
\left\{2-\frac{3}{L(t)
}[W_{0}(t)+3W_{1}(t)] \right\}
= L(t)^{2}     [\rho_m(t)+p_m(t) ] ,
\label{verteq2b}
\end{align} 
with  $L(t)=\sqrt{(y^{0})^{2}-a(t)^{2}(y^{1})^{2}}$.
 Finally, the continuity equation \eqref{continuityeq} under the above 
considerations becomes:
  \begin{equation}
\left[1-\frac{15}{4L(t)}W_{1}(t)\right]\dot{\rho}_m(t) = 
-3\frac{\dot{a}(t)}{a(t)}[\rho_m(t)+p_m(t)][1-5W_{1}(t)]-\frac{5W_{1}(t)\left[
(y^ { 0 } )^ {2}-L(t)^2\right]}{2L(t)^{3}}   \left[ 
\frac{\dot{a}(t)}{a(t)}\right] \rho_m(t).
\label{interactionequation}
 \end{equation}
 As expected and as usual, out of the three equations 
(\ref{fried1b}),(\ref{fried2b}), (\ref{interactionequation})  only two 
are independent, while out of 
the (\ref{verteq1b}),(\ref{verteq2b}), (\ref{interactionequation}) only two 
are independent.  
  
As we observe, in     F-R-S   cosmology we obtain extra 
terms in the 
Friedmann equations, arising from the richer geometrical structure. In 
particular, the  anisotropic torsion terms quantified by the covector   field 
$A_{\mu}$  introduce additional degrees of freedom on the tangent bundle of 
spacetime, which provide the extra contributions in the  Friedmann equations. 
In 
the case where the internal geometrical structure disappears, namely when 
$W_{0}$ and $W_{1}$ become zero, the above equations recover the standard 
Friedmann equations. Additionally, as it was discussed above, in the scenario 
at hand we obtain an interaction between geometry and matter, which is now 
clear 
by the form of (\ref{interactionequation}), and which in the case 
$W_{0}=W_{1}=0$ recovers the  standard  conservation equation too.

We can re-write the Friedmann equations (\ref{fried1b}),(\ref{fried2b}) in 
their standard form 
\begin{align}
\label{standardFr1}
    3H^2 & = \kappa \left( \rho_m + \rho_{DE} \right) \\
    2\dot H & = -\kappa \left( \rho_m + \rho_{DE} + p_m + p_{DE} \right),
\label{standardFr2}
\end{align}
with $H(t)=\dot{a}(t)/a(t)$ the Hubble function, and where we have introduced 
an effective dark energy density and pressure of the form 
\begin{align}
\label{rhoDE}
   \kappa \rho_{DE} & =  \frac{15}{2L} H 
(W_{1}H+\dot{W}_{0}+\dot{W}_{1} 
 )\\
      \kappa p_{DE} & =  - \frac{5}{2L}\left[W_{1}(\dot 
H+H^2) +\dot{W}_{1}H-\ddot{W}_{0}-\ddot{W}_{1} 
       \right],
\label{pDE}
\end{align}
where we have emitted the explicit time-dependence of the various quantities in 
order to make the notation lighter. Thus, the dark-energy equation-of-state 
parameter is defined as 
\begin{equation}
    w_{DE}\equiv\frac{p_{DE}}{\rho_{DE}}.
\label{wDE}
\end{equation}

Additionally, the conservation equation (\ref{interactionequation}) can be 
written as 
  \begin{equation}
 \dot{\rho}_m + 3H (\rho_m+p_m)=Q, 
\label{interactionequation2}
 \end{equation}
 where the interaction term $Q$ is given by
  \begin{equation}
Q\equiv  \frac{15}{4L} W_1\dot{\rho}_m-\frac{5}{2L^3} W_1\left[
(y^ { 0 } )^ {2}-L^2\right]H\rho_m+15 W_1 H.
\label{interactionterm}
 \end{equation}
 Hence, differentiating (\ref{fried1b}) and inserting into (\ref{fried2b}) 
using (\ref{interactionterm}) we also acquire
  \begin{equation}
 \dot{\rho}_{DE} + 3H (\rho_{DE}+p_{DE})=-Q.
\label{interactionequation2}
 \end{equation}
 One can now clearly see that in the scenario of  
    F-R-S  gravity and 
cosmology we obtain an interaction between geometry and matter, and therefore 
an interaction between the effective dark energy and matter sectors. Such an 
interaction is common in Finsler-like cosmologies 
\cite{Kouretsis:2008ha,Ikeda:2019ckp,Papagiannopoulos:2017whb,
Konitopoulos:2021eav,Savvopoulos:2020} and 
it is very interesting since interacting cosmologies  
\cite{Farrar:2003uw,Wang:2006qw,Chen:2008ft,Salvatelli:2014zta, 
Buen-Abad:2017gxg} are known to have many advantages, including solving the 
coincidence problem  \cite{Zimdahl:2001ar,Jamil:2009eb} as well as alleviating 
cosmological tensions \cite{Pan:2019gop,Khyllep:2021wjd}.\\

 \subsection{Specific model }
 
 For completeness, in this subsection we will examine a specific model. As one 
can see from the general Friedmann equations (\ref{fried1b}),(\ref{fried2b}), 
or 
equivalently from  the effective dark-energy sector (\ref{rhoDE}),(\ref{pDE}), 
the appearance of the arbitrary functions $W_0(t)$ and $W_1(t)$, i.e. of 
$A_0(t)$, $A_1(t)$, makes 
the resulting cosmological phenomenology very capable. The only point that one 
should be careful is that the argument of the square root in $L(t)$ should be 
positive, and thus $y^1$ should be suitably smaller than $y^0$.
 
Let us investigate   a specific  example. For simplicity we focus on dust 
matter, 
namely we assume that $p_m=0$. We solve the  equations 
((\ref{fried1b}),  (\ref{verteq1b}) and
(\ref{interactionequation})  
numerically, and as independent variable we 
use 
the redshift $1+z=1/a$ (we set the present scale factor $a_0=1$). Furthermore,  
we introduce the matter and dark energy density parameters,   $\Omega_{m}\equiv 
\kappa\rho_{m}/(3H^2)$ and  
$\Omega_{DE}= \kappa\rho_{DE}/(3H^2)$ respectively. Lastly,
we impose $\Omega_{DE}(z=0)\equiv\Omega_{DE0}\approx0.69$  and 
$\Omega_m(z=0)\equiv\Omega_{m0}\approx0.31$  in agreement with observations
\cite{Planck:2018vyg}.

\begin{figure}[ht]
 \includegraphics[scale=0.45]{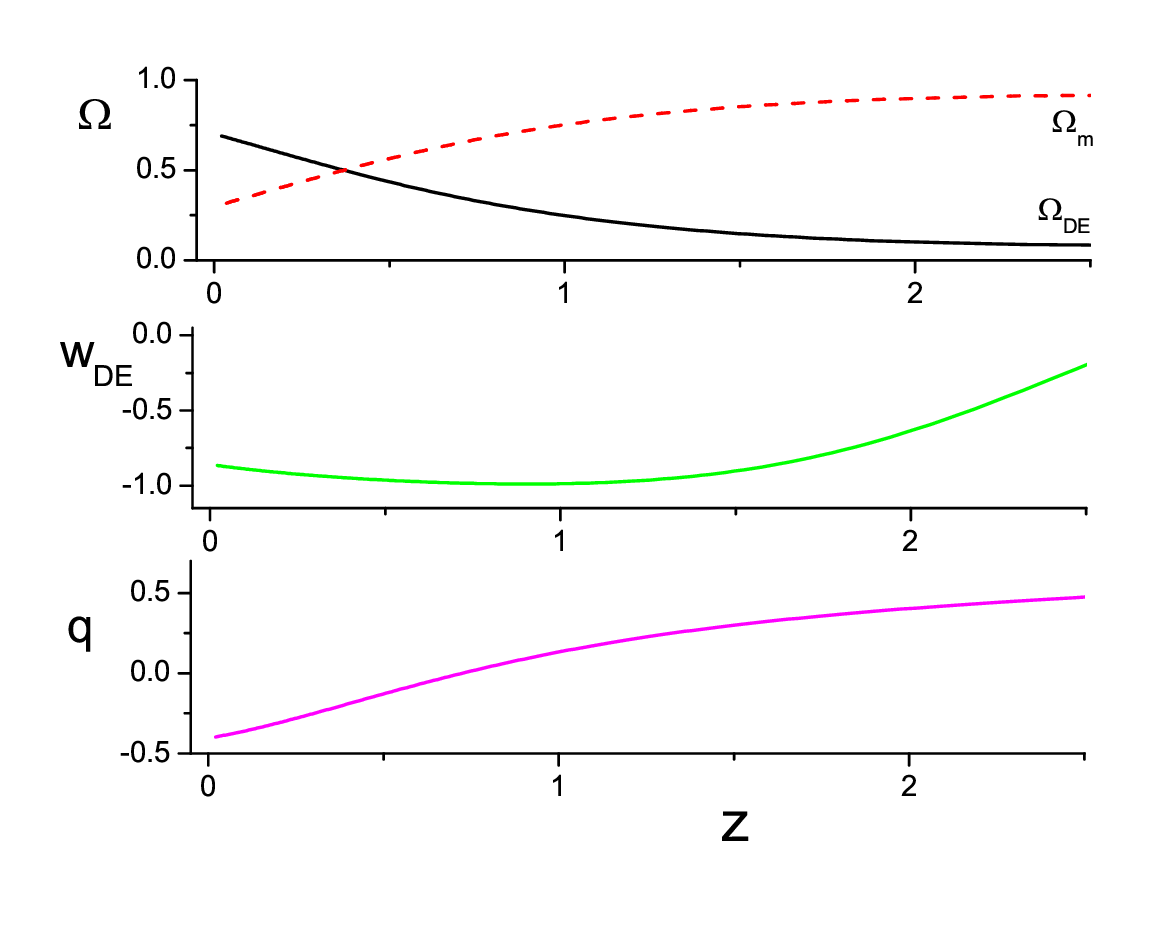}
\caption{
{\it{ {\bf{ Upper graph}}: The effective dark energy    density
parameter $\Omega_{DE}$ (green-solid), and the 
matter   density parameter $\Omega_{m}$   (blue-dashed), as a function of 
the redshift $z$, for      F-R-S  cosmology under     the
ans\"{a}tze 
(\ref{choice1}),(\ref{choice2}), with $y^0=1$, $y^1=10^{-6}$  in units where 
$\kappa=1$. {\bf{Middle graph}}: 
The 
corresponding dark-energy equation-of-state parameter $w_{DE}$. {\bf{Lower 
graph}}: 
The 
corresponding deceleration    parameter $q$.
We have set the initial conditions
$\Omega_{DE}(z=0)\equiv\Omega_{DE0}\approx0.69$ \cite{Planck:2018vyg}.
}} }
\label{Omegas}
\end{figure}

We present the evolution of $\Omega_{m}(z)$ and $\Omega_{DE}(z)$ in the upper 
graph of Fig. \ref{Omegas}. As we can see, we can recover the       universe
thermal history, i.e. the succession of    matter and dark energy 
epochs. Moreover, in the middle graph Fig. \ref{Omegas} we show the evolution 
of 
the corresponding effective dark-energy equation-of-state parameter
$w_{DE}(z)$ according to (\ref{wDE}). For this specific example $w_{DE}$ lies 
in the quintessence regime. Nevertheless, note that according to 
the form of (\ref{rhoDE}),(\ref{pDE}), one could have other scenarios, in which 
$w_{DE}$ can be phantom-like, or experience the phantom divide crossing during 
the evolution. Finally, for completeness, in the lower graph of Fig. 
\ref{Omegas} we depict the 
deceleration parameter $q$, defined as $q=-1-\frac{\dot{H}}{H^2}$. As we can 
see, the transition from acceleration to deceleration happens at 
$z_{tr}\approx0.7$, in agreement with observations.

\section{Discussion and concluding remarks}
\label{Conclusions}

In this work we presented for the first time a Friedmann-like construction in 
the 
framework of an osculating   Finsler-Randers-Sasaki geometry, 
building 
the corresponding 
gravitational theory and applying it at a cosmological setup. In particular, 
we considered a vector field in the metric on the total structure of 
a Lorentz tangent bundle, which depends on the position coordinates. In this 
approach,  the curvatures of horizontal and vertical spaces, 
the extra contributions of torsion, non-linear connection and the vector field,
that depend on $x$ and $y(x)$, provide an 
intrinsic richer geometrical structure, with additional degrees of 
freedom, that lead  to extra terms in the field equations. 
 
Applying these modified field equations at a cosmological setup,  considering
explicit   ans\"{a}tze for the      F-R-S   metric 
functions, we 
extracted 
the generalized Friedmann equations for the horizontal and vertical parts of 
R-S 
spacetime,  in which we have the appearance of extra 
terms, which can be collected to build an effective dark energy sector. Hence, 
in the framework of       F-R-S   geometry and gravity, we 
obtain an 
effective dark energy density and pressure arising from the richer underlying 
structure of the tangent bundle. Additionally, as it is common in Finsler-like 
constructions, we acquire an effective interaction between the matter and the 
geometrical sectors, and in particular of   the extra   
    F-R-S   degree 
of freedom and matter energy density.

We elaborated the generalized Friedmann equations numerically for a specific 
model,  replicating the thermal evolution of the universe, encompassing 
distinct 
matter and dark energy epochs. Our analysis revealed that the dark-energy 
equation-of-state parameter could occupy the quintessence or phantom regime, or 
undergo a phantom-divide crossing during the evolution.

Several crucial steps remain to be comprehensively explored. Firstly, a 
thorough investigation into cosmological applications is imperative, involving 
data confrontation from Type Ia Supernovae (SNIa), Baryon Acoustic Oscillations 
(BAO), and Cosmic Microwave Background (CMB) observations. Additionally, one 
could consider different and more complicated   ans\"{a}tze for the  
    F-R-S   metric functions. Another interesting subject 
is the 
investigation of spherically symmetric and black hole solutions in the theory 
at hand. These essential and intriguing inquiries are reserved for future 
research projects.

\section*{Acknowledgments}
  The authors   acknowledge the contribution of the LISA CosWG, and of   COST 
Actions  CA18108  ``Quantum Gravity Phenomenology in the multi-messenger 
approach''  and  CA21136 ``Addressing observational tensions in cosmology with 
systematics and fundamental physics (CosmoVerse)''.

\begin{appendix}
\section{Action variation and general field equations}
\label{Fieldequations}

In this Appendix we    extract the general field equations, performing 
variation of the action (\ref{Hilbert like action}) on the  
 Finsler-Randers-Sasaki 
geometry. In particular, varying (\ref{Hilbert like action}) we acquire
	\begin{align}
	\Delta K = &\, \int_{\mathcal N} d^8 \mathcal U ( R + S) \Delta 
\sqrt{|\Gd|} 
+ \int_{\mathcal N} d^8 \mathcal U \sqrt{|\Gd|} (\Delta  R + \Delta S) 
+\,  2\kappa\int_{\mathcal N} d^8\mathcal U 
\,\Delta\!\left(\sqrt{|\Gd|}\,\mathcal L_M\right) \label{DK},
	\end{align}
	with
	\begin{align}
	\Delta\sqrt{|\Gd|} = & -\frac{1}{2}\sqrt{|\Gd|}\left( g_{\mu\nu}\Delta 
g^{\mu\nu} + v_{\alpha\beta}\Delta v^{\alpha\beta}\right) \label{DG}\lin
	\Delta R = & \, 2g^{\mu[\kappa}\pdot{\alpha}L^{\nu]}_{\mu\nu} \Delta 
N^\alpha_\kappa + \overline{R}_{\mu\nu}\Delta g^{\mu\nu} + \mathcal D_\kappa 
Z^\kappa \label{DR} \lin
	\Delta S = &\, S_{\alpha\beta} \Delta v^{\alpha\beta} + \mathcal D_\gamma 
B^\gamma \label{DS}.
	\end{align}
	In the above expressions we have defined :
	  $\overline R_{\mu\nu} = R_{(\mu\nu)} + \Omega^\alpha_{\kappa(\mu} 
C^\kappa_{\nu)\alpha}$ and
	\begin{align}
	Z^\kappa = &\, g^{\mu\nu}\Delta L^\kappa_{\mu\nu} - g^{\mu\kappa} \Delta 
L^\nu_{\mu\nu} \nonumber\lin
	= &\, -\D_\nu\Delta g^{\nu\kappa} + g^{\kappa\lambda}g_{\mu\nu}\D_\lambda 
\Delta g^{\mu\nu}  +   2\left( g^{\kappa\mu}C^\lambda_{\lambda\alpha} - 
g^{\kappa\lambda}C^\mu_{\lambda\alpha} \right) \Delta N^\alpha_\mu 
\label{Zdef}\lin
	B^\gamma = &\, v^{\alpha\beta}\Delta C^\gamma_{\alpha\beta} - 
v^{\alpha\gamma}\Delta C^\beta_{\alpha\beta} \nonumber\lin
	= &\, -\D_\alpha \Delta v^{\alpha\gamma} + v^{\gamma\delta}v_{\alpha\beta} 
\D_\delta\Delta v^{\alpha\beta} \label{Bdef}.
	\end{align}
	
	 Applying the Stoke theorem on the Lorentz tangent bundle we obtain
	\begin{align}
	&\int_{\mathcal N} d^8 \mathcal{U}\sqrt{|\Gd|}\,\mathcal D_\kappa Z^\kappa 
= 
  \int_{\mathcal N} d^8 \mathcal{U}\sqrt{|\Gd|}\,\mathcal 
T^\alpha_{\kappa\alpha}Z^\kappa \nonumber\\
	& = \int_{\mathcal N} d^8 \mathcal{U}\sqrt{|\Gd|}\,\D_\nu \left[ \mathcal 
T^\beta_{\kappa\beta} \left( -\Delta g^{\nu\kappa} + 
g^{\nu\kappa}g_{\mu\lambda}\Delta g^{\mu\lambda}\right)\right] \nonumber\\
	& \, -\int_{\mathcal N} d^8 \mathcal{U}\sqrt{|\Gd|}\,\left[ -\D_\nu\mathcal 
T^\beta_{\mu\beta} + g_{\mu\nu}\D^\lambda\mathcal 
T^\beta_{\lambda\beta}\right]\Delta g^{\mu\nu} \nonumber\\
	& \, + 2\int_{\mathcal N} d^8 \mathcal{U}\sqrt{|\Gd|}\,\mathcal 
T^\beta_{\kappa\beta}\left( g^{\kappa\mu}C^\lambda_{\lambda\alpha} - 
g^{\kappa\lambda}C^\mu_{\lambda\alpha}\right)\Delta N^\alpha_\mu
	\end{align}
	\begin{align}
	&\int_{\mathcal N} d^8 \mathcal{U}\sqrt{|\Gd|}\,\mathcal D_\kappa Z^\kappa 
= 
\nonumber\\ 
	& \int_{\mathcal N} d^8 
\mathcal{U}\sqrt{|\Gd|}\,\left(\delta^{(\lambda}_\nu\delta^{\kappa)}_\mu - 
g^{\kappa\lambda}g_{\mu\nu} \right)\left(\mathcal D_\kappa\mathcal 
T^\beta_{\lambda\beta} - \mathcal T^\gamma_{\kappa\gamma}\mathcal 
T^\beta_{\lambda\beta}\right) \Delta g^{\mu\nu} \nonumber\\
	& + \int_{\mathcal N} d^8 
\mathcal{U}\sqrt{|\Gd|}\,4\mathcal{T}^\beta_{\kappa\beta}g^{\kappa[\mu}C^{
\lambda]}_{\lambda\alpha} \Delta N^\alpha_\mu \label{DZ}
	\end{align}\\
	\begin{align}
	&\int_{\mathcal N} d^8\mathcal U \sqrt{|\Gd|}\, \D_\alpha B^\alpha =  
-\int_{\mathcal N} d^8\mathcal U \sqrt{|\Gd|}\,C^\mu_{\mu\beta}B^\beta 
\nonumber\\
	& = -\int_{\mathcal N} d^8\mathcal U \sqrt{|\Gd|}\,\D_\alpha \left[ 
C^\mu_{\mu\beta}\Delta v^{\alpha\beta} - 
v^{\alpha\beta}v_{\gamma\delta}C^\mu_{\mu\beta} \Delta v^{\gamma\delta}\right] 
\nonumber\\ 
	& - \int_{\mathcal N} d^8\mathcal U \sqrt{|\Gd|}\,\left( \D_\alpha 
C^\mu_{\mu\beta} - v^{\gamma\delta}v_{\alpha\beta}\D_\gamma 
C^\mu_{\mu\delta}\right)\Delta v^{\alpha\beta},
	\end{align}
	where we have also used the Leibniz rule. Applying the Stokes theorem again 
and 
eliminating the new boundary terms, we find
	\begin{align}
	& \int_{\mathcal N} d^8\mathcal U \sqrt{|\Gd|}\,\D_\alpha B^\alpha  
 = \int_{\mathcal N} d^8\mathcal U 
\sqrt{|\Gd|}\,\left(v^{\gamma\delta}v_{\alpha\beta} - 
\delta^{(\gamma}_\alpha\delta^{\delta)}_\beta \right)\left(\mathcal D_\gamma 
C^\mu_{\mu\delta} - C^\nu_{\nu\gamma}C^\mu_{\mu\delta} \right) \label{DB}.
	\end{align}

	As a last step, the matter part of the action yields
	\begin{align}
	    & 
\!\!\!\!\!\!\!\!\!\!\!\!\!\!\!\!\!\!\!\!\!\!\!\!\!\!\!\!\!\!\!\!\int_{\mathcal 
N} d^8\mathcal U 
\,\Delta\!\left(\sqrt{|\Gd|}\,\mathcal 
L_M\right)  
	    =   \int_{\mathcal N} d^8\mathcal U 
\sqrt{|\Gd|}\frac{1}{\sqrt{|\Gd|}}\frac{\Delta\!\left(\sqrt{|\Gd|}\,\mathcal 
L_M\right)}{\Delta g^{\mu\nu}}\Delta g^{\mu\nu} \nonumber\\
	    &\ \ \ \ \ \ \ \ \ \ \ \ \ \ \ \ \ \ \ \ \  + \int_{\mathcal N} 
d^8\mathcal U 
\sqrt{|\Gd|}\frac{1}{\sqrt{|\Gd|}}\frac{\Delta\!\left(\sqrt{|\Gd|}\,\mathcal 
L_M\right)}{\Delta v^{\alpha\beta}}\Delta v^{\alpha\beta} \nonumber\\
 &\ \ \ \ \ \ \ \ \ \ \ \ \ \ \ \ \ \ \ \ \ + \int_{\mathcal N} d^8\mathcal U 
\sqrt{|\Gd|}\frac{1}{\sqrt{|\Gd|}}\frac{\Delta\!\left(\sqrt{|\Gd|}\,\mathcal 
L_M\right)}{\Delta N^\alpha_\kappa}\Delta N^\alpha_\kappa \label{EM variation}.
	\end{align}
	
	Finally, combining equations \eqref{DK}-\eqref{Bdef}, \eqref{DZ}, 
\eqref{DB}, \eqref{EM variation} and setting $\Delta K = 0$, we result to the 
field equations, namely
\begin{align}
		& \overline R_{\mu\nu} - \frac{1}{2}({R}+{S})\,{g_{\mu\nu}}  + 
\left(\delta^{(\lambda}_\nu\delta^{\kappa)}_\mu - g^{\kappa\lambda}g_{\mu\nu} 
\right)\left(\mathcal D_\kappa\mathcal T^\beta_{\lambda\beta} - \mathcal 
T^\gamma_{\kappa\gamma}\mathcal T^\beta_{\lambda\beta}\right)  = \kappa 
T_{\mu\nu}  \label{feq1ap}\\
		& S_{\alpha\beta} - \frac{1}{2}({R}+{S})\,{v_{\alpha\beta}} + 
\left(v^{\gamma\delta}v_{\alpha\beta} - 
\delta^{(\gamma}_\alpha\delta^{\delta)}_\beta \right)\left(\mathcal D_\gamma 
C^\mu_{\mu\delta} - C^\nu_{\nu\gamma}C^\mu_{\mu\delta} \right) = \kappa 
Y_{\alpha\beta} \label{feq2ap}\\
		& g^{\mu[\kappa}\pdot{\alpha}L^{\nu]}_{\mu\nu} +  2 \mathcal 
T^\beta_{\mu\beta}g^{\mu[\kappa}C^{\lambda]}_{\lambda\alpha} = \kappa\mathcal 
Z^\kappa_\alpha \label{feq3ap},
	\end{align}
 where we have defined the ``energy-momentum tensors''
	\begin{align}
		T_{\mu\nu} &\equiv - 
\frac{2}{\sqrt{|\Gd|}}\frac{\Delta\left(\sqrt{|\Gd|}\,\mathcal{L}_M\right)}{
\Delta g^{\mu\nu}} = - 
\frac{2}{\sqrt{-g}}\frac{\Delta\left(\sqrt{-g}\,\mathcal{L}_M\right)}{\Delta 
g^{\mu\nu}}\label{em1ap}\\
		Y_{\alpha\beta} &\equiv 
-\frac{2}{\sqrt{|\Gd|}}\frac{\Delta\left(\sqrt{|\Gd|}\,\mathcal{L}_M\right)}{
\Delta v^{\alpha\beta}}  = 
-\frac{2}{\sqrt{-v}}\frac{\Delta\left(\sqrt{-v}\,\mathcal{L}_M\right)}{\Delta 
v^{\alpha\beta}}\label{em2ap}\\
		\mathcal Z^\kappa_\alpha &\equiv 
-\frac{2}{\sqrt{|\Gd|}}\frac{\Delta\left(\sqrt{|\Gd|}\,\mathcal{L}_M\right)}{
\Delta N^\alpha_\kappa} = -2\frac{\Delta\mathcal{L}_M}{\Delta 
N^\alpha_\kappa}\label{em3ap},
	\end{align}
	where  
$\delta^\mu_\nu$ and $ \delta^\alpha_\beta$ are the Kronecker symbols.

\section{Friedmann equations}
\label{friedmannequations}

In this Appendix we show how the general field equations  
\eqref{feq1}-\eqref{feq3}, under the cosmological metric 
(\ref{FRW metric1}) and  (\ref{full bundle metric}) give rise to the Friedmann 
equations on the horizontal and vertical space.
 
Let us start from  the horizontal space. 
First we will calculate the trace of the torsion 
$\mathcal{T}^{\beta}_{\nu\beta}$ that is required inside 
(\ref{bteneq}),\eqref{feq1B}. From the torsion definition    (\ref{torsion}) we 
have that:
\begin{equation}
\mathcal{T}^{\alpha}_{\kappa\alpha}=\pdot{\alpha} 
N_{\kappa}^{\alpha}-L_{\alpha\kappa}^{\alpha}  .  
\end{equation}
If we substitute the non-linear connection    \eqref{Nconnection} and the 
components of $L_{\alpha\kappa}^{\alpha}$ from  \eqref{metric d-connection 2} 
we find:
\begin{align}
\mathcal{T}^{\alpha}_{\kappa\alpha}=&-\frac{1}{2}g^{\alpha\gamma}\delta_{\kappa}
h_{\alpha\gamma}+\frac{1}{2}h_{\gamma\delta}\delta_{\kappa}g^{\gamma\delta}
\Rightarrow \mathcal{T}^{\alpha}_{\kappa\alpha}=-\frac{1}{2}\delta_{\kappa}h,
\end{align}
where $h$ is the trace of the metric $h_{\alpha\beta}$ from  \eqref{hab}.
From the above relation we can see that the torsion tensor is of first order in 
terms of the weak metric $h_{\alpha\beta}$.

As a next step we use \eqref{hab} in order to express the 
$h_{\alpha\beta}$-terms in terms of $A_{\alpha}$ and $L = 
\sqrt{-g_{\alpha\beta}y^{\alpha}(x)y^{\beta}(x)}$. In this way we finally find
 \begin{equation}
\mathcal{T}^{\alpha}_{\kappa\alpha}=-\frac{1}{2}\delta_{\kappa}h=\frac{-5}{4L}
\left[2\partial_{\kappa}A_{
\alpha}y^{\alpha}+A_{\alpha}(2\partial_{\kappa}y^{\alpha}-y^{\delta}g^{
\alpha\beta}\partial_{\kappa}g_{\beta\delta})\right]-\frac{5}{2L^{3}}g_{
\alpha\beta}\partial_{\kappa}y^{\alpha}y^{\beta}A_{\gamma}y^{\gamma} = 
B_{\kappa} + \Lambda_{\kappa}    ,
\end{equation}
where we have set 
\begin{align}
&B_{\kappa}=\frac{-5}{4L}\left[2\partial_{\kappa}A_{\gamma}y^{\gamma}+A_{\gamma}
(2\partial_{\kappa}y^{\gamma}-y^{\alpha}g^{\beta\gamma}\partial_{\nu}g_{
\alpha\beta})\right]\label{bk}\\[7pt]
&\Lambda_{\kappa}=\frac{-5}{2L^{3}}g_{\alpha\beta}\partial_{\kappa}y^{\alpha}y^{
\beta}A_{\gamma}y^{\gamma}\label{lk}.
\end{align}
Additionally, we  calculate the $\delta$-derivatives of the terms 
$B_{\kappa}$ and 
$\Lambda_{\kappa}$ as
\begin{align}
&
\!\!\!\!\!\!\!\!\!\!\!\!\!\!\!
\delta_{\mu}B_{\kappa}=-\frac{5}{4L}\Big[2\partial_{\mu}\partial_{\kappa}A_{
\gamma }
y^{\gamma} + 
\partial_{\kappa}A_{\gamma}(2\partial_{\mu}y^{\gamma}-y^{\delta}g^{
\gamma\epsilon}\partial_{\mu}g_{\delta\epsilon}) + 
2\partial_{\mu}A_{\gamma}\partial_{\kappa}y^{\gamma} + 
2A_{\gamma}\partial_{\mu}\partial_{\kappa}y^{\gamma}
\nonumber \\
& \ \ \  \ \ \  \ \ \ \ \ \ \  -    
\partial_{\mu}(A^{\epsilon}\partial_{\kappa}g_{\gamma\epsilon})y^{\gamma} + 
\frac{1}{2}A^{\epsilon}\partial_{\kappa}g_{\gamma\epsilon}(2\partial_{\mu}y^{
\gamma} - y^{\delta}g^{\beta\gamma}\partial_{\mu}g_{\beta\delta})\Big] 
\nonumber \\
& \ \ \ + 
\frac{1}{L^{3}}\Big(2\partial_{\kappa}A_{\gamma}g_{\alpha\beta}\partial_{\mu}y^
{\alpha}y^{\beta}y^{\gamma} + 
2A_{\gamma}\partial_{\kappa}y^{\gamma}g_{\alpha\beta}\partial_{\mu}y^{\alpha}
y^{\beta} - 
A^{\epsilon}\partial_{\kappa}g_{\gamma\epsilon}g_{\alpha\beta}\partial_{\kappa}
y^{\alpha}y^{\beta}y^{\gamma} \Big),
\end{align}
\begin{align}
&\delta_{\mu}\Lambda_{\kappa} = 
\frac{-5}{2L^{3}}\Big[\partial_{\mu}g_{\alpha\beta}\partial_{\kappa}y^{\alpha}y^
{
\beta}A_{\gamma}y^{\gamma} \!+\! 
g_{\alpha\beta}\partial_{\mu}\partial_{\kappa}y^{\alpha}y^{\beta}A_{\gamma}y^{
\gamma} \!+\! 
g_{\alpha\beta}\partial_{\kappa}y^{\alpha}\partial_{\mu}y^{\beta}A_{\gamma}y^{
\gamma}\! +\!
 g_{\alpha\beta}\partial_{\kappa}y^{\alpha}y^{\beta}\partial_{\mu}A_{\gamma}y^{
\gamma} 
\nonumber\\ 
&
\ \ \ \ \ \ \ \ \ \ \ \ \ \ \ \ \ \ \ \ \ \
- 
\frac{1}{2}\partial_{\mu}g_{\alpha\delta}y^{\delta}\partial_{\kappa}y^{\alpha}A_
{\gamma}y^{\gamma} + 
g_{\alpha\beta}\partial_{\kappa}y^{\alpha}y^{\beta}A_{\gamma}\partial_{\mu}y^{
\gamma} + 
\frac{1}{2}g_{\alpha\beta}\partial_{\mu}g_{\delta\epsilon}\partial_{\kappa}y^{
\alpha}y^{\beta}y^{\delta}g^{\gamma\epsilon}A_{\gamma}\Big]
\nonumber\\ 
&\ \ \ \ \ \ \ \ \ \ \ \ 
+ 
\frac{15}{4L^{5}}\left(-\partial_{\mu}g_{\delta\epsilon}y^{\delta}y^{\epsilon} 
 - 2g_{\delta\epsilon}\partial_{\mu}y^{\delta}y^{\epsilon} + 
2g_{\delta\epsilon}y^{\delta}N^{\epsilon}_{\mu}\right)g_{\alpha\beta}\partial_{
\kappa }
y^{\alpha}y^{\beta}A_{\gamma}y^{\gamma}.
\end{align}

In order to proceed, we have to consider an ansatz for $A_{\gamma}$ and 
$y^{\gamma}$. As we mentioned in (\ref{choice1}),(\ref{choice2}), in this 
work we will focus on the case  
$A_{\gamma}=(A_{0}(t),A_{1}(t),0,0)$ and $ 
y^{\gamma}=(y^{0},y^{1},0,0)$, 
with $y^{0}$, $y^{1}$ constants and $A_{0}(t)$, $A_{1}(t)$ time-dependent 
functions, 
which is consistent with FRW symmetries.
Under these ans\"{a}tze, \eqref{bk},\eqref{lk}  give
\begin{align}
&B_{\nu}=-\frac{5}{4L}( 
2\partial_{\nu}A_{\gamma}y^{\gamma}-A_{\alpha}y^{\gamma}g^{\alpha\epsilon}
\partial_{\nu}g_{
\gamma\epsilon} )\\ 
&\Lambda_{\nu}=0 ,
\end{align}
where $L=\sqrt{(y^{0})^{2}-(y^{1})^{2}a^{2}}$.
Similarly,  \eqref{bteneq} becomes
\begin{eqnarray} 
&&
\!\!\!\!\!\!\!\!\!\!\!\!\!\!\!\!\!\!\!\!\!\!\!\!\!\!\!\!
\mathcal{B}_{\mu\nu}=\frac{-5}{4L}\Big[ 
2y^{\gamma}(\partial_{\mu}\partial_{\nu}A_{\gamma} - 
\Gamma^{\kappa}_{\mu\nu}\partial_{\kappa}A_{\gamma}) - 
2y^{\delta}g^{\gamma\epsilon}\partial_{(\mu}A_{\gamma}\partial_{\nu)}g_{
\delta\epsilon}\nonumber\\
&& + 
A_{\gamma}y^{\delta}(g_{\delta\epsilon}\partial_{\mu}\partial_{\nu}g^{
\gamma\epsilon} + 
\frac{1}{2}\partial_{\nu}g^{\gamma\epsilon}\partial_{\mu}g_{\delta\epsilon} + 
\Gamma^{\kappa}_{\mu\nu}
g^{\gamma\epsilon}\partial_{\kappa}g_{\delta\epsilon}
)\Big]
\label{bteneq1}.
\end{eqnarray}
 Hence, inserting the FRW 
metric (\ref{FRW metric1}), the time-component and the trace of the  
 \eqref{bteneq1} finally gives
\begin{align}
\label{Btteq}
&\mathcal{B}_{tt}=\frac{-5}{2L}\left\{y^{0}\ddot A_{0} + y^{1}\ddot A_{1} - 
2y^{1}\dot A_{1}\left(\frac{\dot{a}}{a}\right) + W_{1}\left[2 
\left(\frac{\dot{a}}{a}\right)^2 - \frac{\ddot a}{a}\right]\right\}\\ 
&\mathcal{B}=\frac{-5}{2L}\left\{ -y^{0}\ddot A_{0} - y^{1}\ddot A_{1} - 
3y^{0}\dot A_{0}\left(\frac{\dot{a}}{a}\right) - y^{1}\dot 
A_{1}\left(\frac{\dot{a}}{a}\right) + W_{1}\left[ 
\left(\frac{\dot{a}}{a}\right)^2 + \frac{\ddot a}{a}\right]\right\},
\label{Btteq2}
\end{align}
with $W_{1}(t)=A_{1}(t)y^{1}$ according to definition (\ref{W1choice1}).

As a last step we take  the  time component of  \eqref{feq1B} and its 
trace, inserting all the above expressions, and substituting the FRW 
metric (\ref{FRW metric1}). After some straightforward manipulations we obtain 
two equations, namely
 \begin{align}
&\left(\frac{\dot{a}}{a}\right)^2 -\frac{\kappa}{3}\rho_m = 
-\frac{1}{3}(\mathcal{B}_{tt} + \mathcal{B})\\ 
&\frac{\ddot{a}}{a}+\frac{\kappa}{6}(\rho_m + 
3p_m)=\frac{1}{3}(\mathcal{B}_{tt}-\frac{1}{2}\mathcal{B}),
 \end{align}
 where $\rho_m$ and $p_m$ are respectively the energy density and pressure of 
the perfect fluid energy-momentum tensor $T_{\mu\nu}$. Lastly, inserting 
(\ref{Btteq}),(\ref{Btteq2}) we obtain 
   \begin{align}
\label{fried2bapp}     
&\left[1-\frac{5}{2L(t)}W_{1}(t)\right]\left[\frac{\dot{a}(t)}{a(t)}\right]^{2} 
- 
\frac{5}{2L(t)}[\dot{W}_{0}(t)+\dot{W}_{1}(t)]\frac{\dot{a}(t)}{a(t)}=\frac{
\kappa } {3} \rho_m(t)\\ 
&\left[1-\frac{5}{4L(t)}W_{1}(t)\right]
\frac{\ddot{a}(t)}{a(t)}+\frac{1}{2}
\left[\frac{ \dot{a} (t)}{a(t) }
\right]^{2}-\frac{5}{2L(t)}\dot{W}_{1}(t)\frac{\dot{a}(t)}{a(t)}+\frac{5}{4L(t)}
[\ddot {W}_{0}(t)
+\ddot{W}_{1}(t)]=-\frac{\kappa}{2}p_m(t).
\label{fried2bappB}
\end{align}

In order to extract the cosmological equations arising from the vertical space 
we begin from   \eqref{vab} and \eqref{hab}.
From those we can calculate :
\begin{equation}
\label{vup}
		v^{\alpha\beta}(x,y(x)) = g^{\alpha\beta}(x) - h^{\alpha\beta}(x,y(x)), 
\end{equation}
where 
\begin{align}
\label{hup}
		h^{\alpha\beta}(x,y(x)) = 
\frac{1}{L}A_{\gamma}(y^{\beta}(x)g^{\alpha\gamma} + 
y^{\gamma}(x)g^{\alpha\beta} + y^{\alpha}(x)g^{\beta\gamma}) + 
\frac{1}{L^3}A_{\gamma}y^{\alpha}(x)y^{\beta}(x)y^{\gamma}(x).
	\end{align}
The vertical energy-momentum tensor is:
\begin{align}
    &Y^{\alpha\beta} = (\rho_m + p_m)y^{\alpha}y^{\beta} + p_mv^{\alpha\beta}  
\Rightarrow\\[7pt]
    &Y^{\alpha\beta} = p_mg^{\alpha\beta} + (\rho_m + p_m)y^{\alpha}y^{\beta}  
- 
\frac{p_m}{L}A_{\gamma}(y^{\beta}(x)g^{\alpha\gamma} + 
y^{\gamma}(x)g^{\alpha\beta} + y^{\alpha}(x)g^{\beta\gamma}) - 
\frac{p_m}{L^3}A_{\gamma}y^{\alpha}(x)y^{\beta}(x)y^{\gamma}(x),
\label{Yup}
\end{align}
 where we used relations  \eqref{vup} and \eqref{hup}.
 
 Hence, starting from the vertical field equation \eqref{feq2b}, by 
raising the indices and substituting   \eqref{Yup}, we obtain:
\begin{align}
-\frac{R}{2\kappa}&\left[g^{\alpha\beta} 
-\frac{1}{L}A_{\gamma}(y^{\beta}g^{\alpha\gamma} +  
y^{\gamma}g^{\alpha\beta} + y^{\alpha}g^{\beta\gamma}) - 
\frac{1}{L^3}A_{\gamma}y^{\alpha}y^{\beta}y^{\gamma}\right] =\nonumber\\
&p_m g^{\alpha\beta} + (\rho_m + p_m)y^{\alpha}y^{\beta} - 
\frac{p_m}{L}A_{\gamma}(y^{\beta}g^{\alpha\gamma} 
+ y^{\gamma}g^{\alpha\beta} + y^{\alpha}g^{\beta\gamma})
-\frac{p_m}{L^3}A_{\gamma}y^{\alpha}y^{\beta}y^{\gamma},
\label{genfried}
\end{align} 
where $R=\frac{\ddot {a}}{a}+\frac{{\dot a}^{2}}{{a}^{2}}$.\\[7pt]
If we take $\alpha=\beta=0$ and $\alpha=\beta=1$ in the above equation 
\eqref{genfried} we finally extract the vertical cosmological equations:
\begin{align}
\label{verteq1}
&\left\{p_m(t)+\frac{1}{2\kappa}\left\{\frac{\ddot{a}(t)}{a 
(t)}+\left[\frac{\dot{a}(t)}{a(t)}\right]^{2} 
\right\}\right\}\left\{1-\frac{1}{L (t)}[3W_{0}(t)+W_{1}(t)]+\frac{1}{L(t)^{3}
}(y^{0})^{2}[W_{0}(t)+W_{1}(t)]\right\}=[\rho_m(t)+p_m(t)](y^{0})^{2},\\[7pt]
& 
  \left\{p_m(t)+\frac{1}{2\kappa}\left\{\frac{\ddot{a}(t)}{a 
(t)}+\left[\frac{\dot{a}(t)}{a(t)}\right]^{2} 
\right\}\right\}
\left\{2-\frac{3}{L(t)
}[W_{0}(t)+3W_{1}(t)] \right\}
= L(t)^{2}     [\rho_m(t)+p_m(t) ] .
\label{verteq2}
\end{align} 

\end{appendix}

\end{document}